\documentclass{svproc}
\usepackage[square,sort,comma,numbers]{natbib}
\usepackage{url}
\usepackage{graphicx}
\newcommand{\lcdm}{$\Lambda$CDM}

\newcommand{\fnl}{f_{\rm NL}}
\newcommand{\ud}{{\rm d}}

\begin{document}
\mainmatter              % start of a contribution
\title{Intensity mapping: a new window into the cosmos}
\titlerunning{Intensity mapping}  % abbreviated title (for running head)
\author{Hamsa Padmanabhan}
\authorrunning{Hamsa Padmanabhan} % abbreviated author list (for running head)
\tocauthor{Hamsa Padmanabhan}
\institute{Canadian Institute for Theoretical Astrophysics,\\
60 St. George Street, \\
Toronto, ON M5S 3 H8, Canada\\
\email{hamsa@cita.utoronto.ca} \footnote{Invited talk at the `Quantum Theory 
and 
Symmetries' (QTS-XI) conference, Montreal, QC, Canada, July 2019.}\\
\texttt{}
}

\maketitle              % typeset the title of the contribution

\begin{abstract}
The technique of intensity mapping (IM) has emerged as a powerful tool to 
explore the universe at $z < 6$. IM measures the integrated emission from 
sources over a broad range of frequencies, unlocking significantly more 
information than traditional galaxy surveys. Astrophysical uncertainties, 
however, constitute an important systematic in our attempts to constrain 
cosmology with IM. I describe an innovative approach which allows us to fully 
utilize our current knowledge of astrophysics in order to develop cosmological 
forecasts from IM. This framework can be used to exploit synergies with other 
complementary surveys, thereby opening up the fascinating possibility of 
constraining physics beyond $\Lambda$CDM from future IM observations.
\end{abstract}
\section{Introduction}

Intensity mapping (IM) has emerged as a novel, powerful probe of cosmology over 
the last decade \citep[e.g.,][]{kovetz2019}. In this technique, the 
aggregate intensity of spectral line emission is mapped out to probe the 
underlying large-scale structure, without resolving individual systems. This 
offers a three-dimensional picture of the formation and growth of baryonic 
material (primarily neutral hydrogen - HI), with the frequency dependence 
tracing their redshift evolution. In contrast to traditional galaxy surveys, 
which reach their sensitivity limits at $z \sim 2$, probing only about a few 
percent\citep{loeb2008} of the comoving volume of the observable universe, IM 
has the potential to  directly access
the exciting `dark ages' of the universe ($z \sim 1000 - 30$) immediately 
following the decoupling 
of 
radiation and matter, the formation and turning on of the first luminous sources 
($z \sim 30-15$), and, ultimately, the \textit{epoch of 
reionization}: the second major phase 
transition of (almost) all the cosmological baryons (believed to have completed 
around $z \sim 6$). IM can also provide valuable insights into astrophysical 
phenomena on smaller scales: probing the interstellar medium (ISM), the site of 
active star formation in most 
normal galaxies, through mapping the carbon monoxide (CO) line emission 
\citep[which acts as a tracer of 
molecular hydrogen, $H_2$, e.g. Ref.][]{hpco} and the 158 $\mu m$ fine stucture 
transition of the [CII] species 
\citep[singly ionized carbon; e.g. Ref. ][]{hpcii}. 

Besides offering exciting astrophysical insights, the unique ability of IM  to 
efficiently probe vast volumes of the universe  makes it an ideal probe of 
cosmology and fundamental physics -- such as modified gravity and dark energy 
models, the nature of dark 
matter, inflationary scenarios and several others. However, due to the
complex interplay between astrophysics and cosmology in IM surveys, this 
requires a precise quantification of the impact of astrophysics on the 
robustness of cosmological constraints. In this article, we summarize recent 
work exploring various facets of this inter-relationship, specifically focussing 
on the effect of  astrophysical uncertainties on the precision and accuracy of 
cosmological forecasts from future IM surveys. We also describe 
how such analyses can open up the fascinating possibility of using IM to constrain
physics beyond the standard model of $\Lambda$CDM.

\section{Forecasts for cosmology and astrophysics}

The challenge for using line-intensity mapping to constrain cosmology and astrophysics is 
twofold: (i) the foregrounds --- both galactic and 
extragalactic  ---
 are orders of magnitude stronger than the signal, and 
(ii) the astrophysics of the tracer itself (such as HI) serves as an effective `systematic' in deriving 
cosmological constraints from intensity maps. The former constraint
may be mitigated using techniques such as \textit{foreground avoidance and 
subtraction}, 
since the frequency structure of the foregrounds are estimated to be very different from those of 
the signal \citep[e.g.,][]{santos2005}. The latter effect, which may be referred to as the `astrophysical 
systematic',
can be effectively addressed by quantifying the 
impact of our uncertainty in the knowledge of the tracer astrophysics, on the observable intensity 
fluctuations. In the case of 21 cm intensity mapping, this can be done by using 
a \textit{data-driven, halo model framework} which uses a parametrized form for 
the HI-halo mass relation $M_{\rm HI} (M,z)$, given by 
\citep{hparaa2016}:
\begin{equation}
M_{\rm HI} (M,z) = \alpha f_{\rm H,c} M 
\left(\frac{M}{10^{11}\,M_{\odot}/h}\right)^{\beta} 
\exp\left[-\left(\frac{v_{{\rm c},0}}{v_{\rm c}(M,z)}\right)^3\right] \,
\end{equation}
with free parameters (i) $\alpha$, the fraction of HI relative to cosmic 
$f_{\rm H,c}$, (ii) $\beta$, the logarithmic slope of the HI-halo mass relation, 
and (iii) $v_{\rm c,0}$, the lower virial velocity cutoff below which haloes 
preferentially do not host HI. Similarly, the small-scale HI density profile, $\rho_{\rm HI}(r,M,z)$ can be 
described by:
\begin{equation}
\rho_{\rm HI} (r;M,z) = \rho_0 \exp\left[-\frac{r}{r_{\rm 
s}(M,z)}\right],\label{rhodefexp}
\end{equation}
with $r_{\rm s}$ defined as $r_{\rm s}(M,z)\equiv R_{\rm v}(M)/c_{\rm HI}(M,z)$ 
and $R_{\rm v}(M)$ denoting the virial radius of the dark matter halo of mass 
$M$. The normalization $\rho_0$ is fixed by requiring the HI mass within the 
virial radius $R_v$ to be equal to $M_{\rm HI}$ at each $(M,z)$. Here, $c_{\rm 
HI} (M,z)$ denotes the concentration parameter defined as \citep{maccio2007}:
\begin{equation}
c_{\rm HI}(M,z) = c_{{\rm HI},0} \left(\frac{M}{10^{11} M_\odot} 
\right)^{-0.109} \frac{4}{(1+z)^\eta}.
\end{equation}
with the free parameters $c_{\rm HI, 0}$ and $\eta$ describing the 
normalization and redshift evolution respectively. Given the combination of the all the data in HI available at present (DLAs, HI galaxy surveys and presently available IM observations), the best-fitting values and uncertainties for the HI astrophysical parameters are constrained to be \citep{hparaa2016, hpar2017}: 
$\alpha = 0.09 \pm 0.01, \ \beta = -0.58 \pm 0.06, \ \log ( v_{{\rm c},0}/ {\rm km  \ s}^{-1}) = 1.56 \pm 0.04, 
\ c_{\rm HI,0} = 28.65 \pm 1.76$ and $\eta = 1.45 \pm 0.04$.

\begin{figure*}
\includegraphics[width = \textwidth]{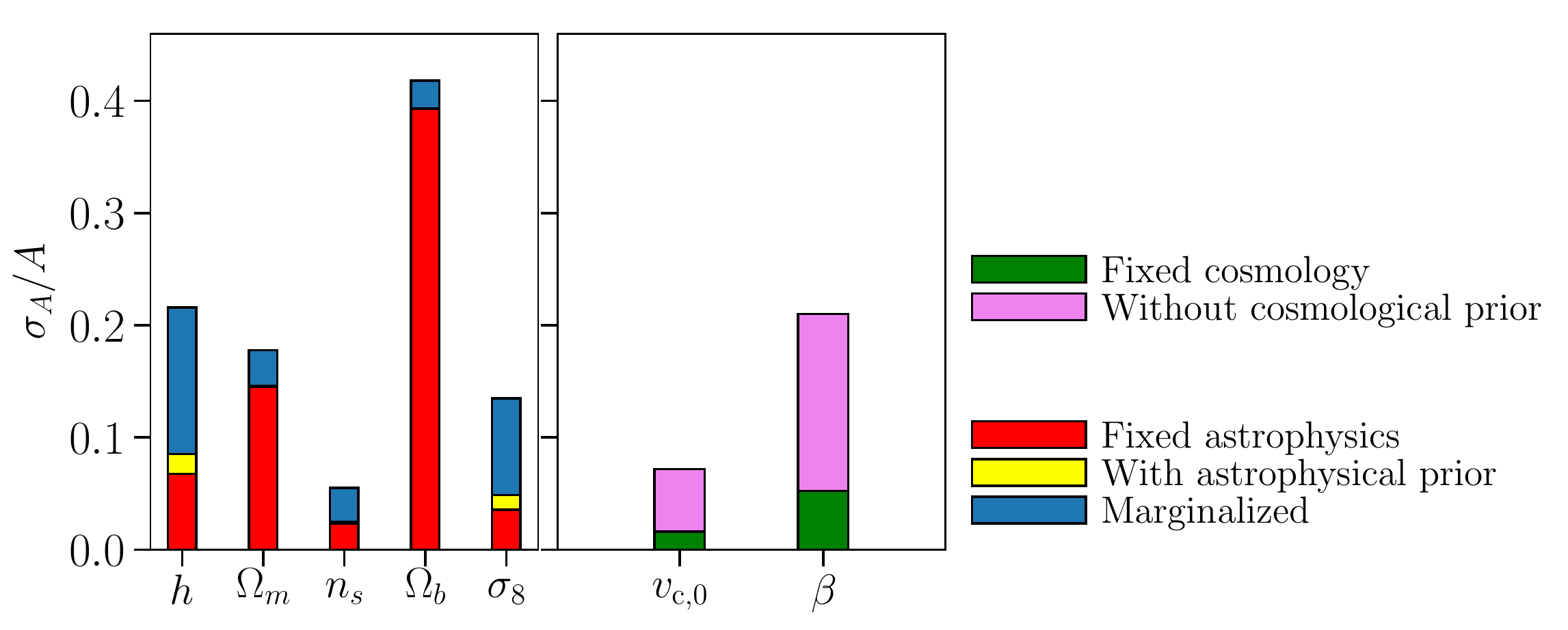} 
\caption{\textit{Left panel:} Constraints on the cosmological 
parameters from a SKA I MID-like configuration, (i) marginalized over 
astrophysics, (ii) with fixed astrophysics, 
and  (iii) with the astrophysical prior coming from the present data. 
\textit{Right panel:} Astrophysical forecasts 
 (i) without cosmological 
priors, and (ii) with fixed cosmology. Figure adapted from Ref. \citep{hparaa2018}.}
\label{fig:skafid}
\end{figure*}

The full, nonlinear power spectrum of HI intensity fluctuations comprises one- and 
two-halo terms, which are given by:
\begin{equation}
P_{\rm HI, 1h}(k,z) = \int \ud M \, n_{\rm h}(M, z) \left[\frac{M_{\rm HI} 
(M)}{\bar{\rho}_{\rm HI}(z)}\right]^2 \ |u_{\rm HI} (k|M,z)|^2.\label{eq:PHI1h}
\end{equation}
and
\begin{equation}
P_{\rm HI, 2h} (k,z) = P_{\rm lin}(k,z)
\times\left[
\int \ud M\,n_{\rm h}(M, z) b_{\rm h}(M, z, k) \frac{M_{\rm HI} 
(M)}{\bar{\rho}_{\rm HI}(z)} |u_{\rm HI} (k|M,z)| \right]^2,\label{eq:PHI2h}
\end{equation}
with 
\begin{equation}
u_{\rm HI}(k|M,z) = \frac{4 \pi}{M_{\rm HI} (M,z)} \int_0^{R_{\rm v}(M)}\ud r\, 
\rho_{\rm HI}(r;M,z) \frac{\sin (kr)}{kr} r^2,
\end{equation}
and 
\begin{equation}
P_{\rm HI}(k,z) = P_{\rm HI, 1 h}(k,z) + P_{\rm HI, 2h}(k,z) 
\end{equation}
is the total HI power spectrum. In the above expressions, the $\bar{\rho}_{\rm HI}(z)$ denotes the average HI density at redshift $z$.
The observable, on-sky quantity in an IM survey is the projected
 \textit{angular power spectrum}, denoted by $C_{\ell}^{\rm HI}(z)$ (for the HI case), which enables a 
tomographic analysis of clustering in multiple redshift bins without the 
assumption of an underlying cosmological model \citep[e.g.,][]{seehars2016}. The expression for $C_{\ell}^{\rm HI}(z)$  can 
be constructed using the Limber approximation \citep[accurate to within 1\% for scales above $\ell \sim 10$; e.g. Ref.][]{limber1953},  using the angular window function, $W_{\rm HI}(z)$ of the survey, as:
\begin{equation}
C_{\ell}^{\rm HI} (z) = \frac{1}{c} \int dz  \frac{{W_{\rm HI}}(z)^2 
H(z)}{R(z)^2} 
P_{\rm HI} [\ell/R(z), z]
\label{cllimber}
\end{equation}
where $H(z)$ is the Hubble parameter at redshift $z$, and $R(z)$ is the 
comoving distance to redshift $z$. 
From the above angular power spectrum, and given an experimental configuration, 
a Fisher forecasting formalism can be used for predicting constraints on the 
various cosmological [$\{h, \Omega_m, n_s, \Omega_b, \sigma_8\}$]  and 
astrophysical [$\{c_{\rm HI}, \alpha, \beta, \gamma, v_{\rm c,0}\}$] 
parameters, generically denoted by $p_{\mu}$. The Fisher matrix element 
corresponding to parameters $\{p_{\mu}, p_{\nu}\}$ at a particular redshift bin 
centred at $z_i$ is given by:
\begin{equation}
F_{\mu\nu} (z_i) = \sum_{\ell < \ell_{\rm max}} \frac{\partial_\mu C_\ell^{\rm 
HI}(z_i)
\partial_\nu C_\ell^{\rm HI}(z_i)}{\left[\Delta C_\ell^{\rm HI}(z_i)\right]^2},
\label{eq:Fisher}
\end{equation}
where $\partial_\mu$ is the partial derivative of $C^{\rm HI}_{\ell}$ with respect to 
parameter $p_\mu$. The standard deviation, $\Delta C_{\ell} (z_i)$ is defined 
in terms of the noise of the experiment, $N_{\ell}^{\rm HI}$ and the sky 
coverage of the survey, $f_{\rm sky}$:
\begin{equation}
\Delta C_\ell^{\rm HI} = \sqrt{\frac{2}{(2 \ell + 1)f_{\rm sky}}} 
\left(C_\ell^{\rm HI} + N_\ell^{\rm HI}\right),\label{eq:Delta_Cl}
\end{equation}
\begin{figure*}
\includegraphics[scale = 0.5]{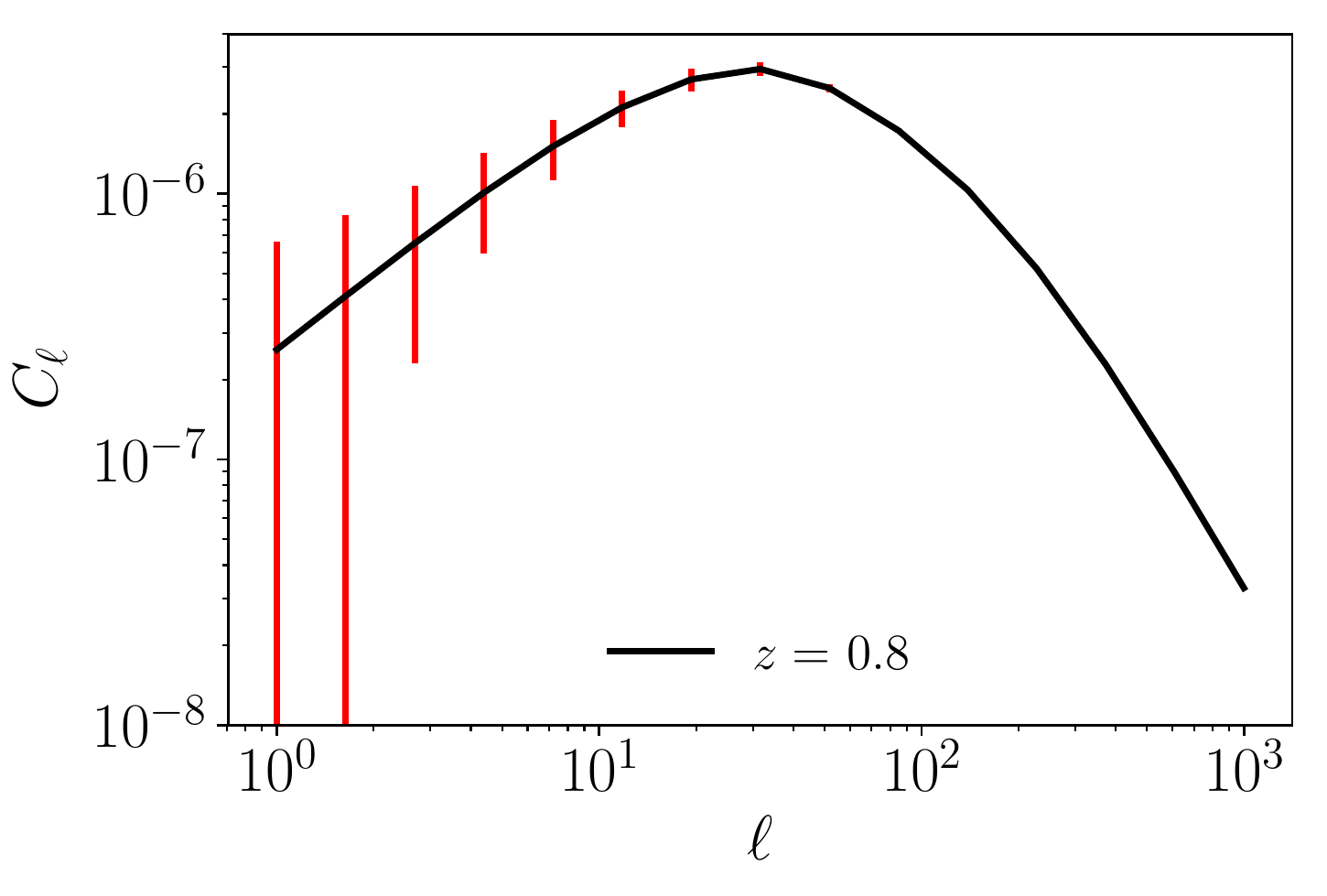}
\caption{Angular power spectrum $C^{\rm HI}_{\ell}$ from  at redshift  0.8, using the 
best-fitting HI astrophysical parameters from the halo model. The error bars 
shown in red represent the standard deviation $\Delta C^{\rm HI}_l$, calculated 
for a SKA 1 MID-like configuration.}
\label{fig:cl}
\end{figure*}
An example angular power spectrum at $z \sim 0.8$ with its associated standard 
deviation (illustrated by the error bars) for a SKA I MID-like experimental 
configuration  is shown in Fig. \ref{fig:cl}. The full Fisher matrix for an 
experiment is constructed by summing over the individual Fisher matrices in 
each of the $z$-bins in the range covered by the survey: $F_{ij, \rm cumul} = 
\sum_{z_i} F_{ij} (z_i)$. Given the cumulative Fisher matrix, we can calculate 
the standard errors on the parameter $p_i$ for the cases of fixing and marginalizing 
over other parameters: $\sigma^2_{i, \rm fixed} = F_{ii}^{-1}$, and $\sigma^2_{i, 
\rm marg} = (F^{-1})_{ii}$. This is useful to quantify the degradation in 
cosmological constraints when astrophysical parameters are marginalized over.
We find that (as shown in Fig. \ref{fig:skafid}) although the astrophysical  
information broadens the cosmological constraints, the broadening is, for the 
large part, mitigated by the prior information coming from the present 
knowledge of the astrophysics, quantified by the halo model.  We also find that 
experiments reaching lower $z$-values achieve more precise cosmological 
constraints, as do those having a larger tomographic coverage \citep[such as 
the SKA I MID; Ref.][]{hparaa2018}. 

\begin{figure*}
\includegraphics[scale = 1, width = 
\textwidth]{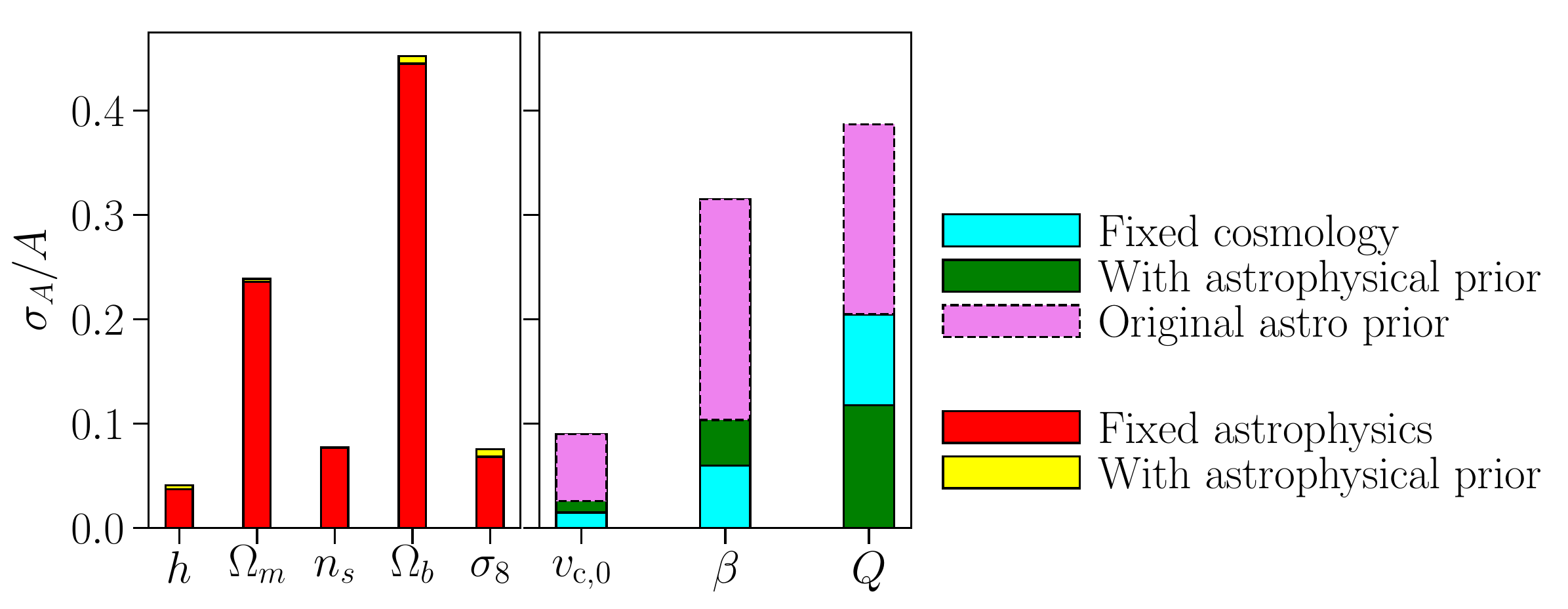}
\caption{Cross-correlation forecasts for a 
CHIME-DESI like survey combination. Fractional errors, $\sigma_A/A$ are plotted 
for  $A = 
\{h, \Omega_m, n_s, \Omega_b, \sigma_8, v_{c,0}, \beta\}$, and an additional 
parameter $Q$ which quantifies the scale dependence of the optical galaxy bias. 
\textit{Left panel:} 
Constraints on the cosmological parameters (i) without marginalizing 
over astrophysics, (ii) with the astrophysical prior coming from the present 
data. \textit{Right panel:}
Constraints on astrophysical parameters when: (i) not 
marginalizing over cosmology and (ii) marginalizing over cosmology and adding 
the astrophysical prior. The violet bands in the right panel show the extent of 
the prior in each case. Figure adapted from Ref. \citep{hparaa2019}.}
\label{fig:forecasts}
\end{figure*}

Cross-correlating 21 cm intensity maps with galaxy surveys (both photometric 
and spectroscopic) offers exciting prospects for constraining astrophysical and 
cosmological parameters. It can be shown \cite{hparaa2019} that 
cross-correlating such surveys covering similar redshift ranges and sky areas, 
significantly improves astrophysical constraints (see Fig. \ref{fig:forecasts} 
for an example of a CHIME-like and DESI-like survey cross-correlated in the 
northern hemisphere).
Further, cross-correlation is a valuable tool to mitigate the effects of 
contaminants and foregrounds, which are expected to be significantly 
uncorrelated between the two surveys \citep[e.g.,][]{foreground12019}.

The Fisher formalism can also be used to quantify how a complementary effect, 
the accuracy of cosmological constraints, is
affected by the astrophysical prior information. This can be done by 
calculating the \textit{relative biases} on cosmological parameters, induced by 
a wrong assumption on the 
astrophysical ones. Such biases 
can be naturally quantified using the `nested likelihoods' 
\citep[e.g.,][]{Heavens2007}  framework in which the parameter space is 
split into two 
subsets: one containing all the parameters of interest and the other, all those 
deemed `nuisance' or systematic for the analysis being carried out. In the 
present case, these two sets represent `cosmological' and `astrophysical' 
parameters, respectively. The bias on a given cosmological parameter $p_a$, 
denoted by $b_{p_a}$, is computed as:
\begin{equation}
b_{p_a} = \delta p_\alpha F_{b\alpha}\left(\mathbf 
F^{-1}\right)_{ab}.\label{eq:bias}
\end{equation}
Here, $\mathbf F^{-1}$ is the full Fisher matrix of astrophysical and 
cosmological parameters, and $F_{b\alpha}$ represents the submatrix mixing 
cosmological and astrophysical parameters. The term $\delta p_\alpha$ denotes 
the vector of shifts in the astrophysical parameters from their fiducial values:
\begin{equation}
\delta p_\alpha=p_\alpha^{\rm fid}-p_\alpha^{\rm true}.\label{eq:shift}
\end{equation}

\begin{figure}
\centering
\includegraphics[scale = 
0.4]{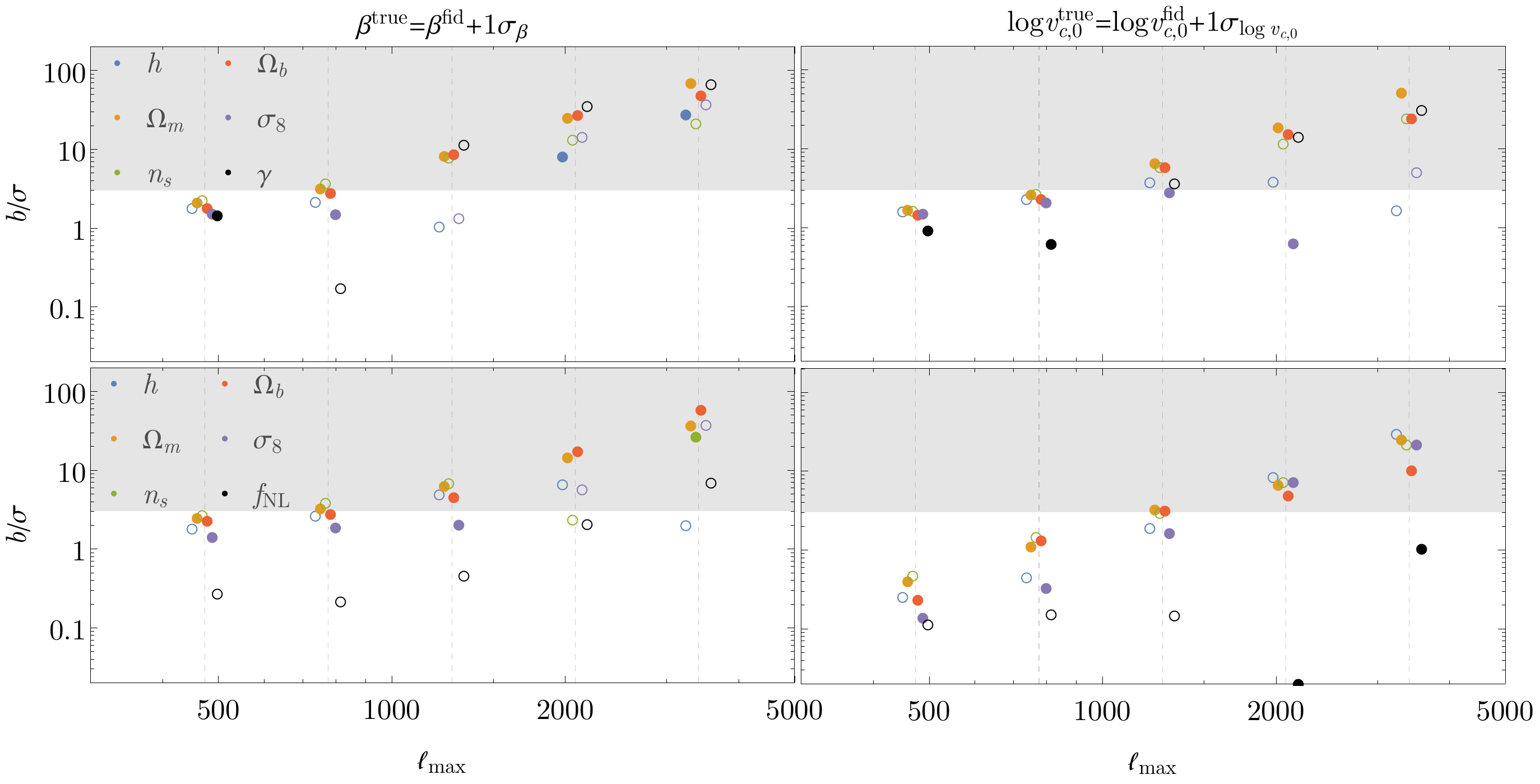}
\caption{Relative bias on cosmological parameters in the extended-$\Lambda$CDM 
framework with a SKA I MID-like experimental configuration, obtained on 
shifting either 
astrophysical parameter, $\beta$ (left panels) or $\log v_{\rm c,0}$ (right panels), 
by $1 \sigma$ from its fiducial value. Top panels contain the parameters in 
\lcdm+$\gamma$, and lower panels contain those in \lcdm+$\fnl$. Empty 
(filled) circles indicate negative (positive) values of  biases. Figure from Ref. \citep{stefanohp2019}.}
\label{fig:relbias1}
\end{figure}

An exciting science case for current and future intensity mapping surveys lies 
in exploring effects beyond the standard model of $\Lambda$ CDM cosmology.
Two widely-studied examples of beyond-$\Lambda$CDM 
physics include (i) the existence of a nonzero $f_{\rm NL}$ parameter that 
quantifies the primordial non-Gaussianity, and (ii) incorporating the effects 
of modified gravity by allowing the growth parameter, $\gamma$ to deviate from 
its fiducial value of 0.55. It can be shown \citep{stefanohp2019} that these 
two effects lead to easily characterizable signatures on the intensity mapping 
power spectrum, by affecting the quantities $n_h(M)$ and $b_h(M,z)$, i.e. the 
abundance and  bias of dark matter haloes. Thus, they can be incorporated in a 
straightforward manner into the angular power spectrum and as such,  the Fisher 
forecasting formalism can be used to compute relative constraints on these 
observables in the presence of astrophysical uncertainties. Fig. 
\ref{fig:relbias1} shows the relative biases on all the standard cosmological 
parameters, as well as $f_{\rm NL}$ and $\gamma$, induced by a deviation of 
either astrophysical parameter, $v_{\rm c,0}$ or $\beta$ from its fiducial 
(i.e. best-fit) value, as a function of the maximum multipole $\ell_{\rm max}$ 
considered in the analysis. The figure reveals that the relative biases on the 
standard cosmological 
parameters and the modified gravity parameter $\gamma$ all remain within a few 
$\sigma$ as long 
as the $\ell$ range stays below 1000. Remarkably, astrophysical uncertainties 
are found to have negligible effects on the $f_{\rm NL}$ parameter, despite 
it being strongly linked to the HI bias.

\section{Conclusions}
In this article, we have explored the ability of current and future intensity 
mapping surveys to provide stringent constraints on cosmology and fundamental 
physics. A data-driven, halo model framework is well-positioned to mitigate the 
`astrophysical systematic' effect on the precision and accuracy of cosmological 
forecasts from these surveys. Such an approach is a powerful tool to test 
extensions to the $\Lambda$CDM framework, such as primordial non-Gaussianity 
and deviations from General Relativity at cosmic scales, as well as to mitigate 
foregrounds through cross-correlating future intensity mapping and optical galaxy 
surveys. In the future, combining these datasets with more traditional probes of the high-redshift universe has the potential to uncover fundamental physics constraints from the hitherto unexplored epochs of Cosmic Dawn and reionization.

\section*{Acknowledgements} I thank the organizers of the Quantum Theory and 
Symmetries (QTS - XI) conference for the invitation to a productive and 
enriching meeting. I thank my collaborators  Adam Amara, Stefano Camera and Alexandre 
Refregier with 
whom most of the work described here was done, and several others for very 
useful and 
interesting discussions.

%

%\begin{thebibliography}{6}

\def\aj{AJ}                   
\def\araa{ARA\&A}             
\def\apj{ApJ}                 
\def\apjl{ApJ}                
\def\apjs{ApJS}               
\def\ao{Appl.Optics}          
\def\apss{Ap\&SS}             
\def\aap{A\&A}                
\def\aapr{A\&A~Rev.}          
\def\aaps{A\&AS}              
\def\azh{AZh}                 
\def\baas{BAAS}
\def\jcap{JCAP}
\def\jrasc{JRASC}             
\def\memras{MmRAS}
\def\na{New Astronomy}
\def\nat{Nature}
\def\mnras{MNRAS}             
\def\pra{Phys.Rev.A}          
\def\prb{Phys.Rev.B}          
\def\prc{Phys.Rev.C}          
\def\prd{Phys.Rev.D}          
\def\prl{Phys.Rev.Lett}       
\def\pasp{PASP}               
\def\pasj{PASJ}
\def\physrep{Phys. Repts.}
\def\qjras{QJRAS}             
\def\skytel{S\&T}             
\def\solphys{Solar~Phys.}     
\def\sovast{Soviet~Ast.}      
\def\ssr{Space~Sci.Rev.}      
\def\zap{ZAp}                 
\let\astap=\aap
\let\apjlett=\apjl
\let\apjsupp=\apjs

\small{
\bibliographystyle{spmpsci}
\bibliography{mybib}
}

%\end{thebibliography}

\end{document}